\newif\ifdraft
\drafttrue

\ifdraft
\documentclass[aps, prb, preprint, superscriptaddress, amssymb, amsmath, floatfix, citeautoscript]{revtex4-1}
\else
\documentclass[aps, prb, reprint, superscriptaddress, amssymb, amsmath, floatfix, citeautoscript]{revtex4-1}
\fi

\usepackage{siunitx}
\usepackage{graphicx}

\begin{document}

\title{Heat conduction in multifunctional nanotrusses studied using Boltzmann transport equation}
\author{Nicholas G. Dou}
\affiliation{Division of Engineering and Applied Science, California Institute of Technology, Pasadena, CA 91125}
\author{Austin J. Minnich}
\affiliation{Division of Engineering and Applied Science, California Institute of Technology, Pasadena, CA 91125}

\begin{abstract}
Materials that possess low density, low thermal conductivity, and high stiffness are desirable for engineering applications, but most materials cannot realize these properties simultaneously due to the coupling between them. Nanotrusses, which consist of hollow nanoscale beams architected into a periodic truss structure, can potentially break these couplings due to their lattice architecture and nanoscale features. In this work, we study heat conduction in the exact nanotruss geometry by solving the frequency-dependent Boltzmann transport equation using a variance-reduced Monte Carlo algorithm. We show that their thermal conductivity can be described with only two parameters, solid fraction and wall thickness. Our simulations predict that nanotrusses can realize unique combinations of mechanical and thermal properties that are challenging to achieve in typical materials.
\end{abstract}

\maketitle


\newcommand{\figoctet}[2][]
{
\begin{figure*}[#1]
\centering
\includegraphics[width=#2]{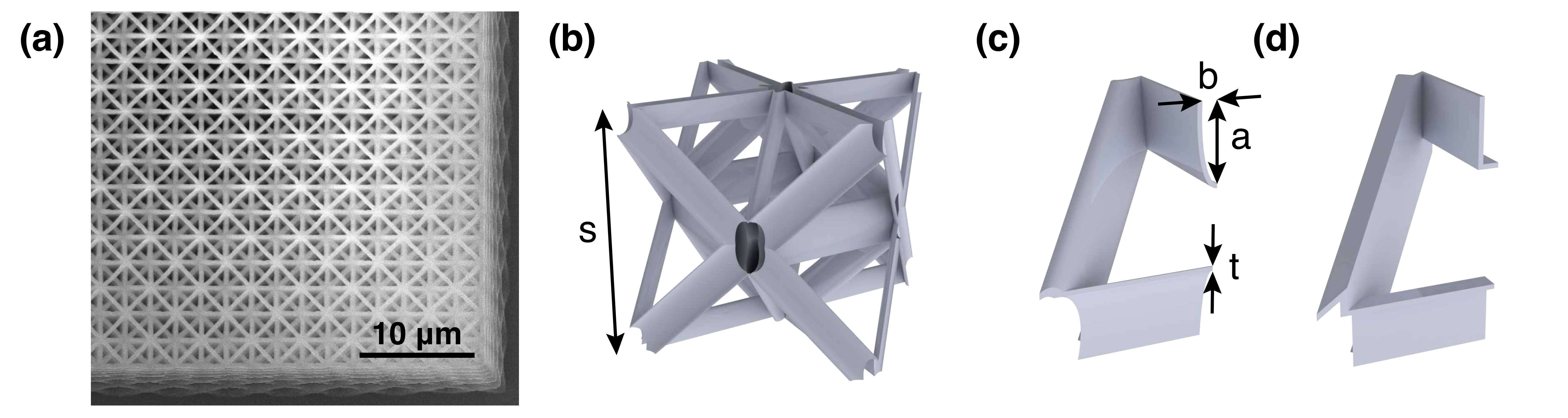}
\caption{(a) Scanning electron microscopy image of a nanotruss showing the hierarchical structure. The octet (b) unit cell and (c) representative subunit, labeled with tunable truss parameters including unit cell size $s$, major axis width $a$, minor axis width $b$, and wall thickness $t$. We use the (d) ``rectangular subunit'' as the simulation geometry, with a spatial structure that is identical to that of the octet.}
\label{fig:octet}
\end{figure*}
}

\newcommand{\figexample}[2][]
{
\begin{figure}[#1]
\centering
\includegraphics[width=#2]{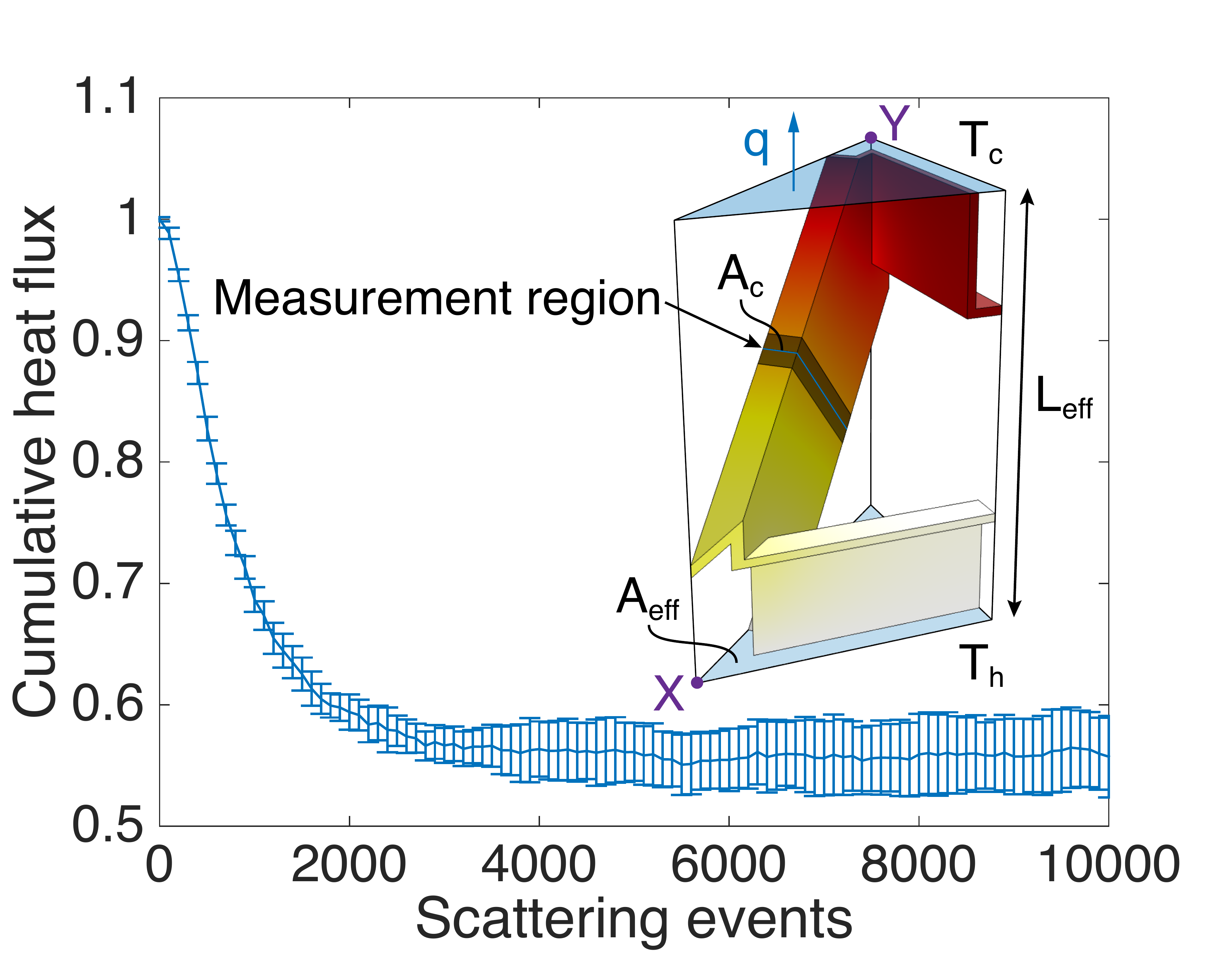}
\caption{Example convergence of average heat flux with respect to scattering events in a MC simulation. Cumulative heat flux is normalized by the first scattering event contribution. The inset shows the FEM temperature field and measurement region with cross-sectional area $A_c$. Thermal conductivity is calculated from Fourier's law using the total heat current $q$, temperature difference $T_\mathrm{h} - T_\mathrm{c}$, effective area $A_\mathrm{eff}$, and effective length $L_\mathrm{eff}$. The full truss structure can be constructed by tessellating the bounding box, which contains two node points $X$ and $Y$ that are used for the thermal resistance model described in the text.}
\label{fig:example}
\end{figure}
}

\newcommand{\figresults}[2][]
{
\begin{figure*}[#1]
\centering
\includegraphics[width=#2]{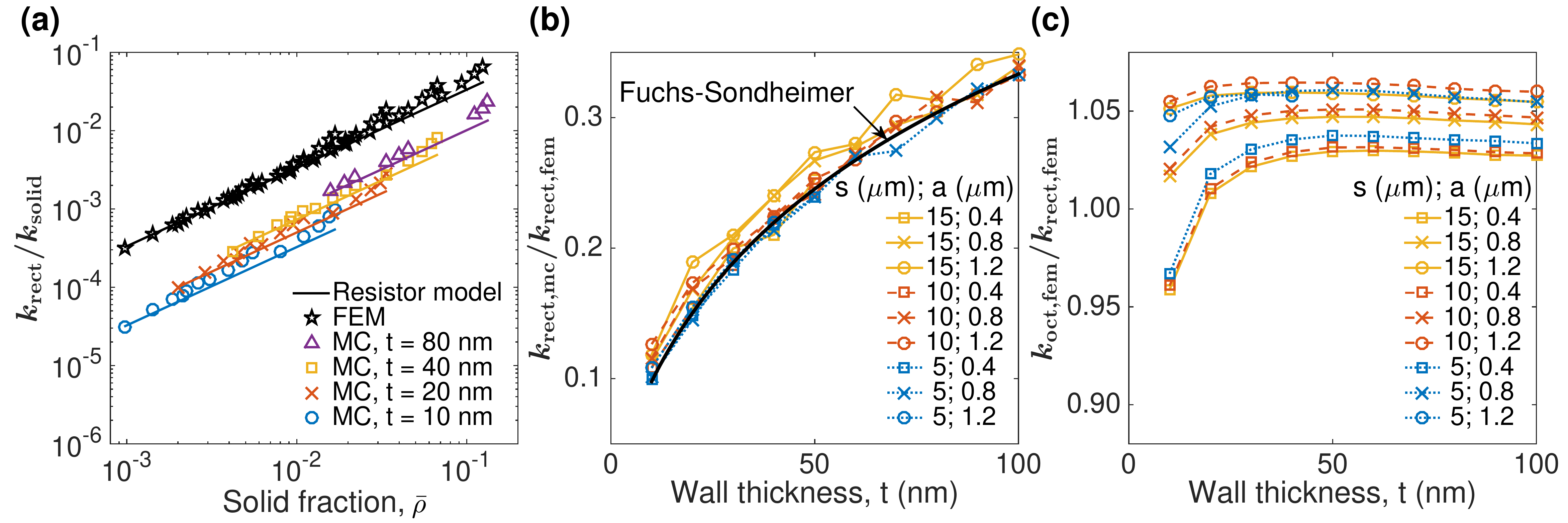}
\caption{(a) MC and FEM thermal conductivities of the rectangular subunit (symbols) versus solid fraction. The thermal conductivities are normalized by the solid phase thermal conductivity. The results are consistent with a simple thermal resistance model (solid lines). (b) Ratio of MC and FEM thermal conductivities (symbols connected by lines) versus wall thickness. Agreement with Fuchs-Sondheimer theory suggests that wall thickness is the critical thermal length. The legend specifies $(s; a)$, where $s$ is the unit cell size and $a$ is the major axis width of each data set. (c) Ratio of octet and rectangular subunit thermal conductivities versus solid fraction obtained using FEM. The ratio is close to unity, thereby justifying the approximation of curved faces as flat faces. }
\label{fig:results}
\end{figure*}
}

\newcommand{\figmatprops}[2][]
{
\begin{figure}[#1]
\centering
\includegraphics[width=#2]{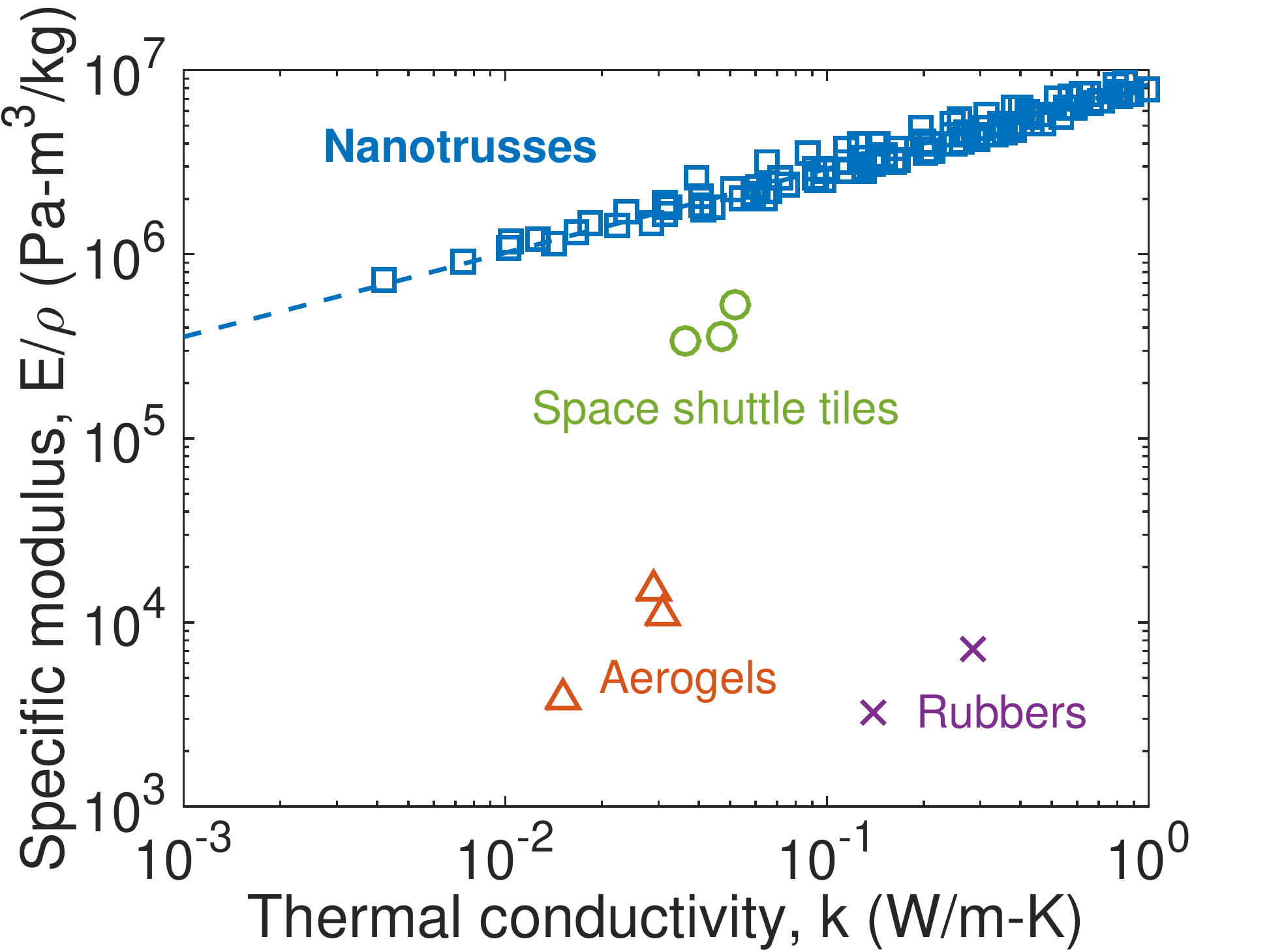}
\caption{Ashby-type plot of specific modulus and thermal conductivity. Silicon octet nanotrusses occupy a previously unreachable region in the parameter space, thus demonstrating their potential in multifunctional applications. The dashed line is a fit to guide the eye.}
\label{fig:matprops}
\end{figure}
}


\ifdraft\else
\figoctet[!]{0.9\textwidth}
\figexample[!b]{0.4\textwidth}
\figresults[!]{\textwidth}
\fi

Multifunctional materials that possess specific combinations of properties in multiple physical domains are highly desirable for engineering applications. For example, the thermal protection system of a spacecraft requires lightweight materials that can withstand large mechanical stresses and thermal gradients. However, despite considerable effort, materials with ultralow density and thermal conductivity yet high elastic modulus are not readily available. The origin of the difficulty is that these physical properties are correlated in common bulk materials. For example, the hardest known material, diamond, also has the highest thermal conductivity \cite{olson_thermal_1993}. Conversely, materials with ultralow thermal conductivity such as graphene and polymer foams \cite{pettes_thermal_2012, wang_modeling_2008}; carbon nanotube films \cite{prasher_turning_2009}, sponges \cite{gui_carbon_2010}, and aerogels \cite{wiener_thermal_2006}; and silica aerogels \cite{zeng_geometric_1995} have poor mechanical properties. A material that combines the desired properties in both the mechanical and thermal domains remains unrealized.

Recent advances in nanofabrication techniques have facilitated the fabrication of ``nanotrusses,'' which are periodic three-dimensional truss structures composed of hollow nanoscale beams, as shown in Fig.~\ref{fig:octet}a. Mechanical studies have established their exceptional stiffness to weight ratio and ability to recover their original shape after 50 percent strain \cite{meza_mechanical_2013}. The hierarchy of length scales that characterizes nanotrusses spans several orders of magnitude---unit cell widths are several microns, truss member diameters are hundreds of nanometers, and wall thicknesses are tens of nanometers. Hierarchical materials can thus outperform bulk materials because the length scales governing various physical processes are decoupled \cite{schaedler_ultralight_2011, seepersad_multifunctional_2005, suralvo_hybrid_2008}. In the case of nanotrusses, mechanical deformation occurs at the scale of truss member lengths, while the length scales relevant for thermal transport are on the order of the wall thickness. The thermal conductivity of nanotrusses is expected to be low due to porosity, and classical size effects can lead to further reduction of thermal conductivity because the wall thickness is comparable to the mean free path of phonons \cite{chen_nanoscale_2005}. However, the thermal transport properties of nanotrusses have never been investigated.

Here, we study heat conduction in the exact nanotruss geometry and evaluate the potential of nanotrusses as a multifunctional material. The challenging task of simulating phonon transport in the complex nanotruss geometry is performed using a variance-reduced MC algorithm to solve the spectral Boltzmann transport equation. We show that nanotruss thermal conductivity can be calculated using only two parameters, solid fraction and wall thickness, which describe geometrical and size effects, respectively. Our result enables the predictive design of nanotruss thermal conductivity.


We begin by describing our simulation approach. The Boltzmann transport equation (BTE) governs phonon transport in the classical size effect regime. The spectral BTE with the relaxation time approximation is given by 
\begin{equation}
\frac{\partial f_\omega}{\partial t} + \mathbf{v}_\omega \cdot \nabla f_\omega
= - \frac{f_\omega - f_\omega^0}{\tau_\omega}.
\end{equation}
where $f_\omega$ is the phonon distribution, $f_\omega^0$ is the local equilibrium distribution, $\mathbf{v}_\omega$ is group velocity, and $\tau_\omega$ is relaxation time. To solve this equation, we employ an efficient variance-reduced Monte Carlo (MC) method developed by P\'{e}raud \textit{et al} \cite{peraud_efficient_2011, peraud_alternative_2012}. This algorithm is around $10^6$ times faster than traditional MC for the problem considered here, and enables simulations of thermal transport in large and complex structures.

The details of the algorithm are provided in Ref.~\citenum{peraud_alternative_2012}. Briefly, the algorithm computes thermal conductivity by following the trajectory of phonon bundles, each carrying a fixed amount of energy. Each of these particles has a frequency, polarization, position, and velocity, which are initialized based on a reference temperature distribution. These properties and state variables are updated as the particle advects and scatters through the domain according to prescribed boundary conditions and internal scattering events. The particle is terminated when the number of scattering events reaches a specified maximum. The average heat flux is calculated using particle effective power, grid cell volume, and cumulative particle displacement through each grid cell.

\ifdraft
\figoctet{\textwidth}
\fi

Here, we focus on nanotrusses with an octet architecture. The complex octet unit cell is reduced into a representative subunit geometry by considering translational and rotational symmetries. Converting curved faces into flat faces while retaining node connectivities leads to the ``rectangular subunit'' in Fig.~\ref{fig:octet}d, the primary simulation geometry in this work. Despite these simplifications, the simulation domain is still very large and complex. Phonons traveling through the structure scatter thousands of times to traverse the micron-scale distance.

Due to the complexity and scale of the domain, code efficiency becomes paramount. The traditional MC algorithm is not optimized for complex geometries. First, the traditional approach for particle advection involves checking every boundary for a collision. For our simulation domain, the computer would need to check 22 planar faces and several edges, because a collision can only occur within the extents of the face. Therefore, boundary checking is quite expensive in this scheme. Secondly, calculating the heat flux field requires defining a grid over the entire domain, which is difficult for a complex geometry. Finally, representing the geometry only as a collection of faces leads to the initialization of particles from boundaries, meaning that the reference temperature distribution is implicitly set to a constant value. Further efficiency improvements can be achieved if the reference temperature reflects the actual temperature distribution more closely \cite{peraud_alternative_2012}.

To overcome these challenges, we decompose the domain into a collection of convex subdomains. In the present case, the simulation domain can be divided into 42 convex prisms and pyramids. This approach offers advantages at every step of the algorithm. First, by tracking which subdomain the particle is in, the algorithm only needs to check the subdomain boundaries for collisions, reducing the number of faces to check from 22 to 6 or fewer. Identifying the next boundary collision or internal scattering location also becomes much simpler compared to the traditional method for non-convex domains. Moreover, verifying whether a point lies inside a convex polyhedron with $N$ faces only requires $N$ dot products, whereas the same task for a non-convex shape may need many more operations. Thus, the subdomain method expedites the advection routine by reducing the number of boundaries that must be checked for collisions, simplifying the determination of whether a collision point is on a boundary face, and eliminating the possibility of particles escaping the domain. 

The subdomain method also avoids complications in setting up a spatial grid in the entire domain by assigning a local coordinate system and grid to each subdomain. The local coordinate system maps each convex polyhedron into a common configuration, such as the unit cube for parallelepipeds. This transformation simplifies the calculation of grid cell occupancy and solid volumes. Each subdomain can have a different grid, which allows computational effort to be concentrated in specific areas of interest.

Perhaps the greatest benefit of splitting the simulation domain into subdomains is the ease of establishing a spatially linear reference temperature, which is done by initializing particles uniformly throughout the domain. First, a random number is drawn to select a subdomain according to relative volumes. Then, three more random numbers are drawn, representing the local coordinates of a point in the chosen subdomain. Using a linear reference temperature reduces computational cost because the actual temperature distribution in the simulation domain is close to linear. Therefore, volumetric generation of particles reduces the variance of the MC algorithm and leads to significantly faster convergence than occurs with the original algorithm.

\ifdraft
\figexample{0.6\textwidth}
\fi

To compute thermal conductivity using this optimized MC method, we establish a vertical temperature gradient across the geometry and calculate the average steady-state heat flux in a measurement region depicted in Fig.~\ref{fig:example} inset. The effective thermal conductivity $k_\mathrm{eff}$ of the corresponding nanotruss is obtained using Fourier's law
\begin{equation}
\frac{q}{A_\mathrm{eff}} = k_\mathrm{eff} \frac{T_\mathrm{h} - T_\mathrm{c}}{L_\mathrm{eff}}
\end{equation}
where $q$ is the total heat current, $T_\mathrm{h} - T_\mathrm{c}$ is the temperature difference, and $L_\mathrm{eff}$ and $A_\mathrm{eff}$ are the bounding box dimensions. In our simulations, we use the following inputs and parameters. Structures have unit cell sizes $s$ from \SIrange{5}{15}{\um}, major axis widths $a$ from \SIrange{0.4}{1.2}{\um}, and wall thicknesses $t$ from \SIrange{10}{100}{\nm}. The minor axis width is $b = a/4$. We take the material of the nanotruss to be silicon with the dispersion and relaxation times given in Ref.~\citenum{minnich_quasiballistic_2011}. In each MC simulation, $10^7$ particles are propagated for $10^4$ scattering events. Figure~\ref{fig:example} demonstrates the MC convergence of heat flux with respect to scattering events for a particular geometry.

Our BTE approach captures thermal conductivity reduction due to both porosity and classical size effects. To isolate the role of size effects, we also solve the heat equation using the finite element method (FEM) software COMSOL and apply the same procedure as above to calculate effective thermal conductivity. The heat equation does not capture phonon size effects, so comparing MC and FEM results is a convenient way to isolate geometrical and size effects in nanotruss structures.


\ifdraft
\figresults{\textwidth}
\fi

We now present the results of our calculations for the octet nanotruss. Figure~\ref{fig:results}a plots MC and FEM thermal conductivities of the rectangular subunit versus solid fraction. The thermal conductivities of all the nanotrusses obtained from FEM collapse onto a single curve, implying that the solid fraction can mostly describe the geometrical factors affecting heat conduction. On the other hand, the MC results lie on separate curves that shift downwards as wall thickness decreases. This decrease in thermal conductivity suggests that size effects are occurring at the length scale of wall thickness, but also that the geometrical reduction in thermal conductivity remains the same as that in the absence of size effects.

To explain the FEM data, we use a thermal resistance network consisting of a single resistor connecting node points $X$ and $Y$ in the subunit bounding box, as indicated in Fig.~\ref{fig:example} inset. The model assumes that no heat flows through the horizontal beams connecting isothermal nodes and that beam lengths are much greater than beam widths ($s \gg a,b,t$), such that the resistance is
\begin{equation}
R = \frac{L_\mathrm{eff}}{k_\mathrm{eff} A_\mathrm{eff}}
= \frac{L_{XY}}{k_\mathrm{solid} A_c}
\end{equation}
where $k_\mathrm{solid}$ is the bulk thermal conductivity of silicon, $L_{XY}$ is the distance between $X$ and $Y$, and $A_c$ is the beam cross-sectional area labeled in Fig.~\ref{fig:example} inset. By relating all geometrical parameters to $s$ and $A_c$, we obtain a simple linear relation between thermal conductivity and solid fraction $\bar\rho$,
\begin{equation}
\frac{k_\mathrm{eff}}{k_\mathrm{solid}} = \frac{32\sqrt{2} A_c}{s^2}
\approx \frac{\bar\rho}{3}
\end{equation}
As shown in Fig.~\ref{fig:results}a, the FEM simulation results are consistent with the theory, particularly for small values of $\bar\rho$ where the thin beam approximation is well satisfied. Therefore, the thermal conductivity of nanotrusses in the absence of classical size effects can be determined using only the solid fraction.

We examine the role of size effects by computing the ratio of MC to FEM thermal conductivities in Fig.~\ref{fig:results}b. When plotted against wall thickness, the data matches well with Fuchs-Sondheimer theory \cite{sondheimer_mean_1952}, which describes thermal transport along thin films. The decrease in thermal conductivity with decreasing wall thickness is caused by boundary scattering at length scales commensurate to phonon mean free paths. The theory explains our simulation results for the entire range of cell sizes and major axes considered in this study, implying that the dominant thermal length scale for nanotrusses in this regime is wall thickness. Further, our calculation shows that the effect of boundary scattering can be described using the well-known analytical equation from Fuchs-Sondheimer theory.

The MC data can be explained by combining the thermal resistance model with Fuchs-Sondheimer theory. The wall thickness dependent family of curves plotted in Fig.~\ref{fig:results}a demonstrates that this hybrid theory can provide a reasonable estimate for the MC thermal conductivities. Hence, these simulation results show that octet nanotrusses can be modeled as a simple thermal resistance network with reduced solid thermal conductivity. Specifically, their thermal conductivity can be determined by two parameters---solid fraction describes geometrical effects through a thermal resistance model, and wall thickness describes size effects through Fuchs-Sondheimer theory. 

To justify the approximation of curved faces as flat faces in this work, Fig.~\ref{fig:results}c compares FEM simulations of the octet and rectangular subunit structures, which have curved and flat faces, respectively. The thermal conductivities are within 6 percent of each other, which validates the similarity of the two geometries. Although these heat equation calculations do not account for classical size effects, we have demonstrated that size effects are primarily dependent on wall thickness, which is the same for both structures. 


\ifdraft
\figmatprops{0.6\textwidth}
\else
\figmatprops[!]{0.4\textwidth}
\fi

Using the calculated thermal conductivities along with an elastic modulus scaling from literature $E \propto (\bar\rho)^{1.61}$ \cite{meza_strong_2014}, we compare the thermomechanical properties of silicon octet nanotrusses to those of other materials. We use an Ashby-type plot of specific modulus versus thermal conductivity, shown in Fig.~\ref{fig:matprops}, to visualize the parameter space. Compared to existing insulation materials such as aerogels and space shuttle tiles, nanotrusses can potentially achieve specific stiffnesses that are 10 to 1000 times higher while reaching even lower thermal conductivities. 

It is possible to further reduce thermal conductivity and increase elastic stiffness by manipulating composition. Nanotrusses constructed with materials such as alumina may have better multifunctional properties than silicon due to lower bulk thermal conductivity and higher specific modulus. In addition, nanotrusses with multilayered walls could achieve further reductions in thermal conductivity by increasing phonon interfacial scattering.

In summary, we have studied thermal transport in multifunctional nanotrusses using a variance-reduced Monte Carlo algorithm to solve the BTE. The thermal conductivity of nanotrusses can be described with two parameters, solid fraction and wall thickness. All of the geometries studied in this work can be fabricated using two-photon lithography and standard nanofabrication techniques. Nanotrusses therefore have considerable potential to fulfill the need for simultaneously lightweight, stiff, and thermally insulating materials in aerospace applications and other structural thermal insulation applications.

\section*{Acknowledgements}

This work was supported by the Air Force Office of Scientific Research (AFOSR) Multifunctional Materials program under grant number FA9550-14-1-0266. The authors thank Lucas Meza and Julia R. Greer for useful discussions.

\bibliography{montecarlo_paper}
\bibliographystyle{unsrt}


\end{document}